\documentclass[11pt]{article}
\usepackage[a4paper,margin=2.4cm]{geometry}
\usepackage{amsmath,amssymb,bm}
\usepackage{graphicx}
\usepackage{booktabs,array}
\usepackage[numbers,sort&compress]{natbib}
\usepackage{caption}
\usepackage[hidelinks]{hyperref}
\usepackage{setspace}
\usepackage{lineno}
\usepackage[section]{placeins}
\graphicspath{{figures/}}
\newcommand{\z}{\bm{z}}

\title{Temporal and spatial couplings in the dynamics of human walking and running}
\author{M. Mert Ankaral\i\\[0.2em]\small ORCID: 0000-0002-1235-5373\\[0.3em]
\small Department of Electrical and Electronics Engineering, Middle East Technical University, Ankara, T\"urkiye\\
\small Center for Robotics and Artificial Intelligence (METU-ROMER), Middle East Technical University, Ankara, T\"urkiye\\
\small \texttt{mertan@metu.edu.tr}}
\date{}

\begin{document}
\maketitle

\begin{abstract}
\noindent Rhythmic behaviours such as walking and running are commonly modelled as dynamical systems around a stable limit
cycle. To keep such models tractable, it is usually assumed that each step depends only on the previous one, that the
limit cycle is fixed, and that the sagittal and frontal planes can be studied separately. Here, we test these
assumptions in seven public datasets of human gait (71 participants), including 16 people who both walked and ran. We
fit linear step-to-step maps and validate them by cross-validation, holding out more data where more data exist. At
first sight, gait has memory: the state two steps back lowers the prediction error by up to 29\%. Yet most of this
temporal coupling vanishes once the operating point may drift slowly and measurement noise is modelled; gait remains
limit-cycle-like, around an operating point that moves. Running keeps more memory than walking in the same people.
Spatial coupling between the planes is present in most participants, mainly from sagittal to frontal, but weak. This
supports planar models of healthy gait, while a repeatable, task-dependent coupling index may serve as a measure of
gait quality. The simple models are wrong, but useful enough.
\end{abstract}


\section{Introduction}

Many biological behaviours, such as locomotion and breathing, are rhythmic and can be modelled as rhythmic dynamical
systems. Every such model neglects some characteristics of the real system as a modelling convenience. Human gait is
usually modelled step by step, as if each step depended only on the one before, around a fixed periodic pattern, and with
the sagittal and frontal planes treated separately. None of these simplifications is exactly true. ``Essentially, all
models are wrong, but some are useful'', wrote George Box \cite{box1987empirical}. The practical questions are how wrong
these simplifications are, and whether they are still useful.

Rhythmic locomotion is commonly described as motion near a stable periodic orbit, a limit cycle
\cite{full1999templates,holmes2006dynamics}. Sampling the state once per step on a Poincar\'e section turns the continuous
dynamics into a step-to-step map, whose fixed point is the intersection of the limit cycle with the section. Near a
hyperbolic fixed point the linearised map describes the local behaviour \cite{hartman1960lemma}, and its eigenvalues, the
Floquet multipliers, are a standard measure of gait stability
\cite{hurmuzlu1994measurement,dingwell2007differences,bruijn2014assessing}. Linear step maps can also be identified directly
from data \cite{revzen2012finding,revzen2015data,maus2015constructing,ankarali2015walking}. Such a model has two core
elements: the limit cycle, which describes the steady-state gait, and the dynamics around it, which describe how the gait
deviates from the limit cycle and returns to it. This paper is about the latter.

Such analyses rest on assumptions that are rarely tested. First, the section state is assumed to be Markov: the state at one
step is enough to predict the next. Second, the operating point is assumed constant, so a single mean is removed from the
sections. Third, the sagittal and frontal planes are often analysed separately. Estimates of gait stability also depend on the
section, on the state variables and on the amount of data \cite{riva2013orbital,riva2013influence,jin2021validation,bruijn2009statistical}.

Each assumption can fail for a known reason. Step-to-step fluctuations contain slow components
\cite{hausdorff1995walking}, step length follows slow changes of walking speed \cite{collins2013two}, and humans regulate
task variables while letting other combinations wander \cite{dingwell2010humans,dingwell2015identifying}. A recent model of
such regulation is only semistable, so the operating point is expected to drift \cite{patil2024model}. Measurement noise is
another reason: a Markov process observed with noise is no longer Markov, and a finite autoregressive model fitted to it
needs extra lags \cite{granger1976time,staudenmayer2005measurement}. Finally, the measured state may miss variables that
matter. Foot placement follows the state of the centre of mass \cite{hof2008extrapolated,wang2014stepping,bruijn2018control},
and in running a spring-mass state is not enough to predict the next step \cite{maus2015constructing,seethapathi2019step}.
Any of these produces apparent memory in the section state.

The planes are also linked. Lateral balance needs active control through foot placement
\cite{kuo1999stabilization,bauby2000active,donelan2004mechanical,oconnor2009direction}, the same muscles contribute to
fore--aft and lateral foot placement \cite{roelker2019muscle}, and fore--aft and lateral margins of stability co-vary
\cite{buurke2021relationship}. In earlier work we found that a single mirrored step map describes left and right steps of
walking well enough for the amount of data available \cite{ankarali2015walking}. That study used sagittal states and one
dataset.

As in our earlier work on symmetry \cite{ankarali2015walking}, we ask two related questions about each simplification:
how wrong is it, statistically, and how useful is the simpler model, in terms of prediction and interpretation? We study
two kinds of coupling in these step-to-step dynamics. \emph{Temporal coupling} is the dependence of a step on
earlier steps, beyond the last one (memory). \emph{Spatial coupling} is the dependence between the sagittal and the frontal
plane from one step to the next. We ask three questions with seven public datasets of walking and running, one of which
records the same people walking and running. Is the section state Markov? If not, is the
memory due to the operating point, to measurement noise, or to a hidden state? Are the sagittal and frontal planes coupled
from one step to the next, and in which direction? We fit linear step maps of
increasing order and validate them by blocked cross-validation with a parsimony rule, holding out more data where more
data are available. We test the constant operating point and relax it to one that drifts slowly. We find that most of the apparent
memory is explained by a drifting operating point and measurement noise, that running keeps more memory than walking within
the same people, and that a further state in running is supported only by the longest recordings. The planes are coupled,
mainly from the sagittal to the frontal plane, but weakly; a coupling index built on this term is repeatable and changes
with the task. A first-order, planar step map around a slowly moving operating point is therefore wrong in a strict sense,
but it is a useful model of healthy gait.

\section{Methods}

\subsection{Data, state and sections}
Seven public datasets were used (table~\ref{tab:data}). Lorenz \& van den Bogert \cite{lorenz2024comprehensive,lorenz2023data}
recorded treadmill walking at 1.2~m\,s$^{-1}$ with rare, brief belt-speed pulses; steps within 2.5~s after a pulse were
excluded. De Graaf et al.\ \cite{degraaf2019influence,degraaf2019data} recorded treadmill walking at 1.25~m\,s$^{-1}$ in seven
arm-swing conditions with Xsens MVN inertial motion capture. Lee et al.\ \cite{lee2026adaptive,lee2026data} recorded
treadmill walking on five slopes with markers; the three 5-min level trials per participant were used. Marotta et al.\
\cite{marotta2021towards,marotta2021data} recorded overground running on a 400-m track with Xsens MVN. Maus et al.\
\cite{maus2015constructing,maus2015data} recorded treadmill running at a self-selected speed with markers. Seethapathi \&
Srinivasan \cite{seethapathi2019step,seethapathi2019data} recorded treadmill running at 2.5, 2.7 and 2.9~m\,s$^{-1}$ with
marker clusters, one recording per speed; the three speeds of a runner were treated as that runner's trials. The OSF dataset \cite{krumm2026multimodal,osf2026data} recorded 18 adults on
an instrumented dual-belt treadmill, walking at 4 and 6~km\,h$^{-1}$ and running at 9~km\,h$^{-1}$ in three interleaved
3-min bouts each, with markers (Plug-in Gait); 16 participants had joint angles. Walking (4~km\,h$^{-1}$) and running
(9~km\,h$^{-1}$) were analysed as two sets of the same people; the fast walk is reported in the text. A further dataset \cite{moore2015elaborate,moore2014data}, with only
about one minute of unperturbed walking per condition, served as a small-data check (electronic supplementary material, section S4).

Raw recordings were reduced to tables of the kinematic state at every heel strike and toe-off, and at 20 equally spaced
phases of every stride, with labels, quality flags and metadata. All analyses use these tables.

\begin{table}[!htbp]
\centering\small
\caption{Datasets. Sections: heel-strike sections in valid step sequences.}
\label{tab:data}
\setlength{\tabcolsep}{4pt}
\resizebox{\linewidth}{!}{%
\begin{tabular}{@{}l >{\raggedright\arraybackslash}p{5.2cm} l r r r@{}}
\toprule
Dataset & Behaviour & Sensors, rate & Participants & Sections & Steps each \\
\midrule
Lorenz 2024 \cite{lorenz2024comprehensive,lorenz2023data}   & treadmill walking, 1.2 m\,s$^{-1}$, rare belt pulses & markers, 100 Hz & 10 & 14\,396 & $\approx$1\,440 \\
de Graaf 2019 \cite{degraaf2019influence,degraaf2019data}    & treadmill walking, 1.25 m\,s$^{-1}$, 7 arm-swing conditions & Xsens, 120 Hz & 12 & 15\,554 & $\approx$1\,300 \\
Lee 2026 \cite{lee2026adaptive,lee2026data}              & treadmill walking, level trials of a slope study & markers, 100 Hz & 13 & 20\,066 & $\approx$1\,540 \\
Marotta 2021 \cite{marotta2021towards,marotta2021data}       & overground running, 400-m track & Xsens, 240 Hz & 8 & 39\,000 & $\approx$4\,900 \\
Maus 2015 \cite{maus2015constructing,maus2015data}        & treadmill running, self-selected speed & markers, 250 Hz & 4 & 14\,066 & $\approx$3\,500 \\
Seethapathi 2019 \cite{seethapathi2019step,seethapathi2019data}  & treadmill running, 2.5--2.9 m\,s$^{-1}$ & marker clusters, 100 Hz & 8 & 13\,124 & $\approx$1\,640 \\
OSF 2026 \cite{krumm2026multimodal,osf2026data} walk           & treadmill walking, 4 km\,h$^{-1}$, same people as OSF run & markers, 250 Hz & 16 & 12\,944 & $\approx$810 \\
OSF 2026 \cite{krumm2026multimodal,osf2026data} run            & treadmill running, 9 km\,h$^{-1}$ & markers, 250 Hz & 16 & 20\,382 & $\approx$1\,270 \\
\bottomrule
\end{tabular}}
\end{table}

\begin{figure}[!htbp]
\centering
\includegraphics[width=\linewidth]{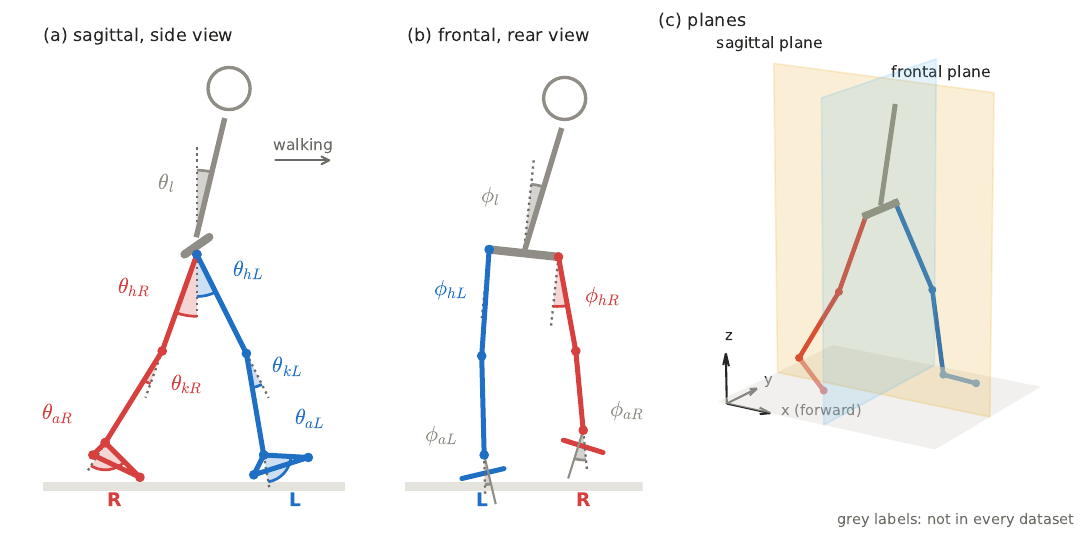}\\[0.3em]
\includegraphics[width=0.78\linewidth]{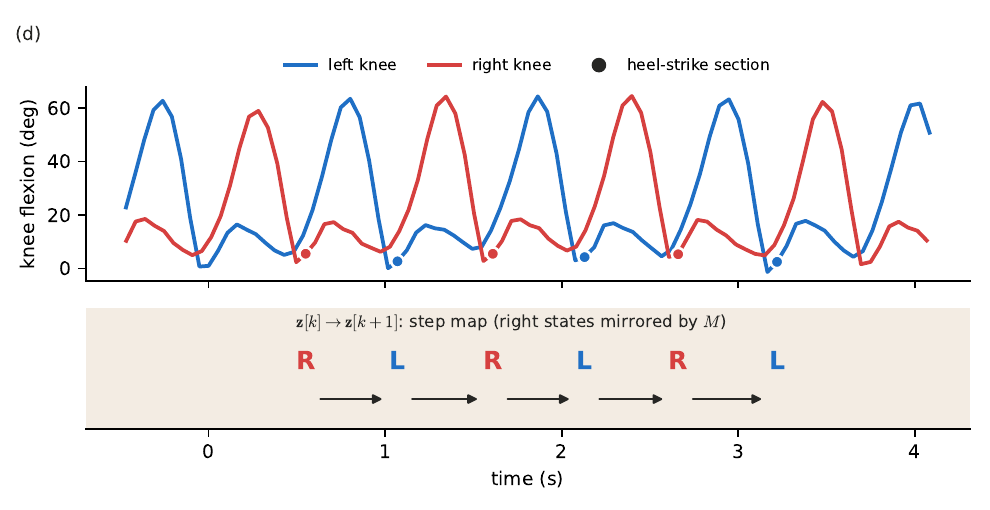}
\caption{State and sections. (\textit{a},\textit{b}) Joint angles of the state for both legs (left: blue; right: red),
each the angle between neighbouring segments (dotted: proximal segment continued; shaded: the angle). Sagittal (side view):
lumbar flexion ($\theta_l$, trunk relative to pelvis), hip, knee and ankle flexion. Frontal (rear view): lumbar lateral
flexion ($\phi_l$), hip ab/adduction and ankle inversion/eversion. Grey labels: not available in every dataset. Each angle
enters the state together with its rate, so the sagittal state has $d_s=14$ variables (12 without the lumbar angle) and
the frontal state 4--10. (\textit{c}) The sagittal plane contains the walking direction and the vertical; the frontal
plane is perpendicular to it. (\textit{d}) Knee flexion of both legs over four strides of one Lorenz recording (degrees).
Dots: heel-strike sections. The step map sends the state at one heel strike to the next (left to right,
right to left); right-strike states are mirrored into left coordinates, so one map describes all steps.}
\label{fig:state}
\end{figure}

The state is the vector of joint angles and their rates (figure~\ref{fig:state}\textit{a}--\textit{c}). Sagittal plane: lumbar flexion and hip,
knee and ankle flexion of both legs ($d_s=14$ with rates; 12 in Lee and Seethapathi, which have no trunk markers). Frontal
plane: lumbar lateral flexion and hip ab/adduction of both legs, and the ankle where the sensors allow it ($d_f=6$ for Lorenz
and Maus, 10 for Xsens and OSF, 8 for Lee, 4 for Seethapathi). Global pelvis angles were not used.
Frontal knee angles were not used because, from markers, they mostly reflect cross-talk from knee flexion
\cite{leardini2005human}. For the marker datasets, angles are planar angles between segment lines, low-pass filtered at
10~Hz (zero phase), and rates are obtained by differentiation and the same filter. For Xsens the joint angles of the MVN model
were used, filtered at 15~Hz. Xsens sagittal angles agree well with optical motion capture; frontal and transverse angles
agree less well \cite{zhang2013concurrent,alamri2018inertial,robertlachaine2017validation}. For Lee the authors' Visual3D
joint angles and for OSF the authors' Plug-in Gait angles (spine angle as the lumbar joint) were used (15~Hz). Seethapathi provides marker clusters without a static trial; hip and knee angles were taken
as relative rotations of the cluster frames from the trial-mean posture, and the ankle angle as the difference of foot and
shank pitch (the feet carry only two to four markers, so no frontal ankle angle).

Heel strikes were taken from treadmill force (Lorenz), from total vertical force with the side assigned by the lower foot
(Maus), from foot impact acceleration (Marotta), from belt forces (OSF: per belt in walking; in running from the total force, with the side given by the more
flexed hip), and from foot position relative to the pelvis \cite{zeni2008two} (de Graaf, Lee, Seethapathi).
Let $\bm q[k]$ be the state at the $k$-th heel strike, alternating left and right. Right-strike states were mirrored into
left coordinates (legs swapped, frontal angles negated), as in \cite{ankarali2015walking}, so one map describes all steps. A
step sequence is a run of at least eight alternating strikes without gaps. Each variable was scaled to unit variance per
participant. All analyses use heel-strike sections (figure~\ref{fig:state}\textit{d}); electronic supplementary
material, section S6, repeats the main tests at 20 phases of the stride.

\subsection{Models}
Write the centred state as $\z[k]=(\z^s[k],\z^f[k])$. For an output plane $p$ with other plane $\bar p$ the candidate maps are
\begin{align}
  \text{AR1:}    &\quad \z^p[k] = A\,\z^p[k-1] + \bm e[k], \notag\\
  \text{AR1+X1:} &\quad \z^p[k] = A\,\z^p[k-1] + C\,\z^{\bar p}[k-1] + \bm e[k], \notag\\
  \text{AR2:}    &\quad \z^p[k] = A\,\z^p[k-1] + B\,\z^p[k-2] + \bm e[k], \label{eq:ladder}\\
  \text{AR2+X1:} &\quad \z^p[k] = A\,\z^p[k-1] + C\,\z^{\bar p}[k-1] + B\,\z^p[k-2] + \bm e[k], \notag\\
  \text{AR3:}    &\quad \z^p[k] = A\,\z^p[k-1] + B\,\z^p[k-2] + D\,\z^p[k-3] + \bm e[k]. \notag
\end{align}
The gain of a model is the relative change of its held-out error with respect to a simpler model. ``Second step'' refers to
AR1$\to$AR2 and ``other plane'' to AR1$\to$AR1+X1. A full-state map $\z[k]=W\z[k-1]+\bm e[k]$ was also fitted, together with
reduced-rank versions $W_r$ \cite{izenman1975reduced} whose rank was chosen by cross-validation.

Eigenvalues of a least-squares step map are biased toward zero by measurement noise and toward one by a drifting operating
point, and the reduced rank counts predictable rather than slow directions. The slow dynamics were therefore estimated with
a latent linear model, $\bm x[n+1]=A\bm x[n]+\bm w[n]$, $\bm y[n]=C\bm x[n]+\bm v[n]$, identified by stochastic subspace
identification with canonical-variate weighting \cite{larimore1990canonical,vanoverschee1996subspace}. The past (three
sections) and the future (four sections) are related through their cross-covariance, which white measurement noise does
not enter; $A$ follows from the shift structure of the estimated observability matrix. The order was chosen by blocked
cross-validation of four-step-ahead predictions with the parsimony rule (held-out share as below, counting only the
training windows that do not overlap a test block). The order starts at 1, as the step maps start at AR1: every
participant shows a one-step dependence, so a model without state is not a candidate. The slow modes are the eigenvalues of $A$. Each
participant was analysed with the full state after per-trial centring, and after removing a minute-scale trend (cubic
B-spline in step index, one knot per 100 steps).

In the basic analysis one mean per participant and side was removed. Two alternatives were compared. In \emph{per-trial
centring} one mean per trial and side was removed. In the \emph{drift model} the offset varies within a trial,
\begin{equation}
  \z[k] = A\,\z[k-1] + \bm c_{\pi(k)}(t_k) + \bm e[k], \qquad \bm c_\pi(t)=\sum_j \bm w_{\pi j} B_j(t),
\end{equation}
with $t_k$ the time of strike $k$, $\pi(k)$ its side, and $B_j$ cubic B-splines with $0$ to $4$ interior knots per minute.
The step map $A$ was not penalised; the spline weights carried a first-difference penalty with weight 10
\cite{eilers1996flexible}, so a constant offset is nested in every spline model. The fixed point implied by the fitted map is
$\bm x_L(t)=(I-A^2)^{-1}(A\bm c_R(t)+\bm c_L(t))$ for the two-step (left$\to$right$\to$left) map.

To test whether measurement noise explains the memory beyond AR1, the model (AR1+noise)
\begin{equation}
  \bm x[k+1] = A\bm x[k] + \bm w[k],\qquad \bm y[k] = \bm x[k] + \bm v[k], \qquad
  \bm w\sim\mathcal N(0,Q),\ \bm v\sim\mathcal N(0,R),\ R\ \text{diagonal},
  \label{eq:ssm}
\end{equation}
was fitted to each participant and output plane by expectation--maximisation (EM) \cite{shumway1982approach,ghahramani1996parameter}.
The observation matrix is the identity, so the latent state has the units of the measured state; $Q$ is a full and $R$ a
diagonal covariance. EM alternates a Kalman smoother (E-step) with closed-form updates of $A$, $Q$ and $R$ (M-step), starting
from the least-squares AR1 map, 70\% of its residual covariance as $Q$ and 15\% of each variable's variance as $R$ (30
iterations; 10 more from the full-data fit within each cross-validation split). The two noises are identifiable because
they enter the autocovariance differently: $R$ adds to the variance only, whereas $Q$ is propagated by $A$ to all lags.
Under this model $\bm y$ is an ARMA(1,1) process \cite{granger1976time}, so an AR2 map can partly mimic it. Predictions are
one-step Kalman predictions. Its parameter count, $d^2+d(d+1)/2+d$, lies between those of AR1 and AR2.

\subsection{Validation and model selection}
Two validation schemes were used. \emph{Population shrinkage with leave-one-trial-out testing} shrinks each participant's map
toward the map fitted on the other participants,
\begin{equation}
  \hat B_s(\lambda) = \arg\min_B \|Y_s-X_sB\|_F^2+\lambda\|B-\hat B_{\text{pop},-s}\|_F^2,
\end{equation}
in the spirit of \cite{james1961estimation,efron1975data,hoerl1970ridge}, and predicts each trial from the participant's
other trials. One $\lambda$ per dataset was chosen on AR1 from $\{0,3,10,\dots,3000\}$. Choosing it without the test
participant (nested cross-validation) changed it for 2 of 97 participants and left every reported gain unchanged (ESM S1).

\emph{Blocked Monte-Carlo cross-validation} (MCCV) fits each participant's own data. Step sequences were cut into blocks of
20 consecutive steps; a share $f$ of the blocks was held out at random, the models were fitted on the rest and scored on
the held-out blocks, and this was repeated 60 times (40 times in the drift test of \S3.4 and 8 times for the state-space
model, which is slower to fit). Blocking limits leakage between dependent samples
\cite{burman1994cross,bergmeir2012use,roberts2017cross}.

The held-out share acts like the penalty of an information criterion: a small share keeps larger models, and a share that
grows with the data, while the training part still grows, makes the choice consistent \cite{shao1993linear,arlot2010survey}.
A fixed share therefore treats 800 and 5\,000 steps alike. We used one share per dataset, the same for all its
participants, kept enough data for fitting and held out the rest,
\begin{equation}
  n_c=\max\bigl(n^{3/4},\;10\,p_{\max}\bigr),\qquad f=1-n_c/n,\qquad 0.2\le f\le 0.9,
  \label{eq:holdout}
\end{equation}
with $n$ the median number of steps per participant in the dataset (per recording in the drift test; units too short to
analyse not counted) and $p_{\max}$
the number of unpenalised regressors per equation of the largest candidate. $n^{3/4}$ is Shao's training size
\cite{shao1993linear}; ten rows per regressor is the usual floor for least squares. The resulting shares ran from 20\% (short
recordings, large models) to 88\% (Marotta; electronic supplementary material, table S1). Fixed shares of 20, 50 and 80\% were kept as a sensitivity check. Among the candidates, ordered by number of parameters, the chosen model is the
simplest one whose mean error is within one standard error of the best model \cite{breiman1984classification,hastie2009elements}, the standard error being that of the paired difference over repetitions. The same rule chose the
number of drift knots and the rank. For the state-space model, each held-out block was predicted by restarting the Kalman
filter at the block start; the first three steps of every block were not scored for any model.

\emph{Coupling index.} The strength of the cross-plane term was summarised per participant and output plane as
\begin{equation}
  \mathrm{CI}=100\,\bigl(1-e_{\text{AR1+X1}}/e_{\text{AR1}}\bigr)\ \%,
  \label{eq:ci}
\end{equation}
the relative held-out error gain from adding the other plane's state at the previous step to the own-plane AR1 map
(blocked MCCV with the dataset share, per-trial and per-side means removed). Both maps were shrunk toward population maps
fitted on the other participants ($\lambda=100$), since individual maps without shrinkage were dominated by estimation
error. $\mathrm{CI}_{\text{fro}}$ measures the sagittal$\to$frontal and $\mathrm{CI}_{\text{sag}}$ the
frontal$\to$sagittal coupling. Repeatability was assessed by computing CI on the first and the second half of each
participant's steps (intraclass correlation ICC(3,1) within dataset; smallest detectable change
$\mathrm{SDC}=1.96\sqrt2\,\mathrm{SEM}$), and task dependence within the same people across the OSF speeds and gaits and
the five Lee slopes.

Paired comparisons used sign tests or Wilcoxon signed-rank tests.

\section{Results}

\subsection{Apparent memory and weak coupling with one operating point}
\begin{figure}[!htbp]
\centering
\includegraphics[width=0.9\linewidth]{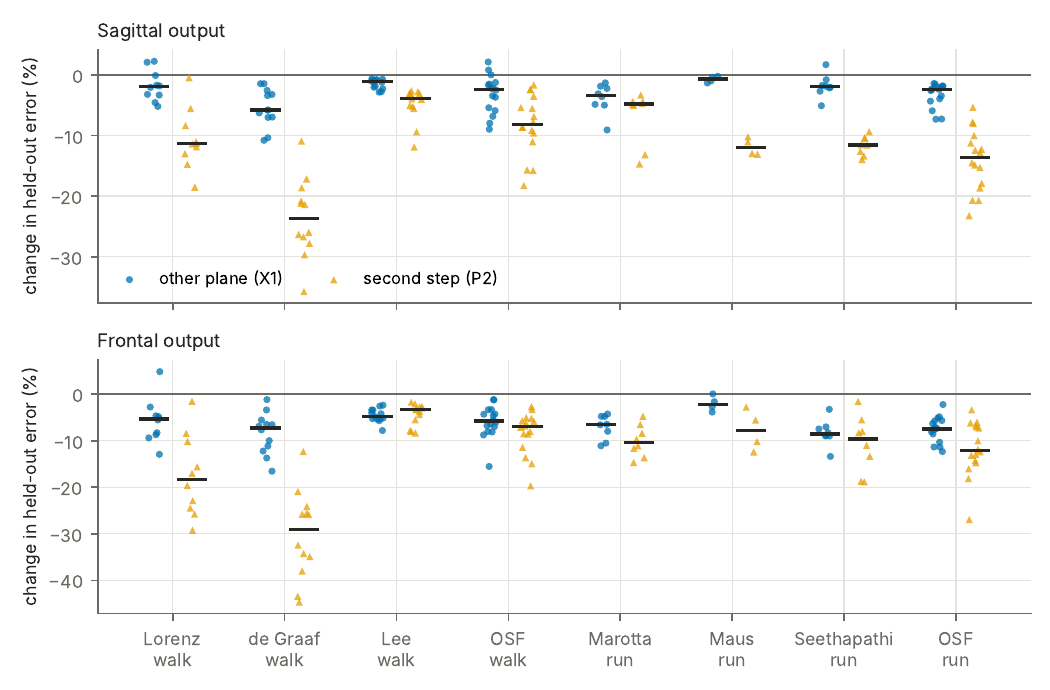}
\caption{What the cross-plane term and the second past step add to the one-step (AR1) map, with one operating point
per participant. Each point is one participant; bars are medians. Vertical axis: change in held-out error,
$100\,(e_{\text{larger}}/e_{\text{AR1}}-1)$ in \%; negative values mean the added term improves the prediction. Blue
circles: adding the other plane's state at the previous step (AR1$\to$AR1+X1). Orange triangles: adding the own plane's
state two steps back (AR1$\to$AR2). Validation: each trial is predicted from the participant's other trials (leave one
trial out), with the map shrunk toward the population map of the other participants. Held-out error is the mean squared one-step prediction error of the output variables, each scaled to unit variance per participant, so it is dimensionless: 1 means no better than the participant's mean section, 0 a perfect prediction. Top: sagittal output;
bottom: frontal output.}
\label{fig:ladder}
\end{figure}

With one operating point per participant, adding the state two steps back lowered the held-out error in every dataset and in
every participant (table~\ref{tab:ladder}, figure~\ref{fig:ladder}). The gain ranged from 3.3\% (Lee, frontal) to 29\%
(de Graaf, frontal). A third step did not help, except slightly in Maus (sagittal, $-0.9$\%). Adding the other plane at the
previous step lowered the error by 0.7--9\%, more for the frontal than for the sagittal output. Once the second step was
included, the cross-plane gain dropped to 0.4--5\%. These gains are relative to the whole section variance, of which
at most a third is predictable from the past (ESM S8); the Discussion asks what they mean.

\begin{table}[!htbp]
\centering\small
\caption{Median change in held-out error (leave-one-trial-out, population shrinkage) when a block is added to the step map;
participants improved in brackets. sag / fro: output plane.}
\label{tab:ladder}
\setlength{\tabcolsep}{4pt}
\resizebox{\linewidth}{!}{%
\begin{tabular}{@{}l c c c c c c c c@{}}
\toprule
 & \multicolumn{4}{c}{Walking} & \multicolumn{4}{c@{}}{Running} \\
\cmidrule(lr){2-5}\cmidrule(l){6-9}
 & Lorenz & de Graaf & Lee & OSF & Marotta & Maus & Seethapathi & OSF \\
\midrule
Other plane (AR1$\to$AR1+X1), sag & $-2.0$\% (8/10) & $-5.8$\% (12/12) & $-1.1$\% (13/13) & $-2.4$\% (14/16) & $-3.4$\% (8/8) & $-0.7$\% (4/4) & $-2.0$\% (7/8) & $-2.4$\% (16/16) \\
\phantom{Other plane (AR1$\to$AR1+X1),} fro & $-5.4$\% (9/10) & $-7.3$\% (12/12) & $-4.9$\% (13/13) & $-5.8$\% (16/16) & $-6.5$\% (8/8) & $-2.2$\% (3/4) & $-8.6$\% (8/8) & $-7.5$\% (16/16) \\
Second step (AR1$\to$AR2), sag & $-11$\% (10/10) & $-24$\% (12/12) & $-4.0$\% (13/13) & $-8.2$\% (16/16) & $-4.8$\% (8/8) & $-12$\% (4/4) & $-12$\% (8/8) & $-14$\% (16/16) \\
\phantom{Second step (AR1$\to$AR2),} fro & $-18$\% (10/10) & $-29$\% (12/12) & $-3.3$\% (13/13) & $-7.0$\% (16/16) & $-10$\% (8/8) & $-7.9$\% (4/4) & $-9.7$\% (8/8) & $-12$\% (16/16) \\
Other plane given two steps, sag & $-0.8$\% (8/10) & $-1.4$\% (12/12) & $-0.7$\% (13/13) & $-1.4$\% (12/16) & $-2.4$\% (8/8) & $-0.4$\% (4/4) & $-0.9$\% (7/8) & $-1.5$\% (14/16) \\
\phantom{Other plane given two steps,} fro & $-2.6$\% (9/10) & $-1.6$\% (11/12) & $-3.6$\% (13/13) & $-2.5$\% (16/16) & $-4.0$\% (8/8) & $-1.1$\% (4/4) & $-5.4$\% (8/8) & $-3.9$\% (15/16) \\
Third step (AR2+X1$\to$AR3), sag / fro & $+0.3$ / $+2.7$\% & $+0.6$ / $+0.8$\% & $+1.1$ / $+3.5$\% & $+1.8$ / $+2.7$\% & $+1.3$ / $+3.7$\% & $-0.9$ / $-0.3$\% & $-0.5$ / $+4.3$\% & $+0.2$ / $+3.4$\% \\
\bottomrule
\end{tabular}}
\end{table}

\begin{figure}[!htbp]
\centering
\includegraphics[width=0.95\linewidth]{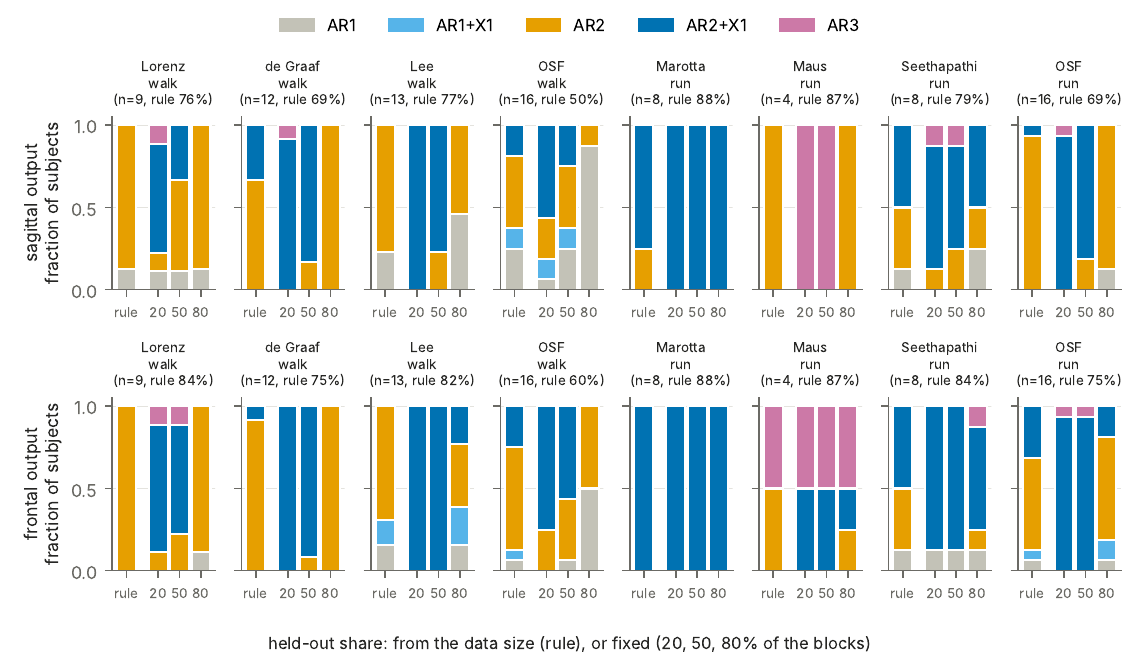}
\caption{Which model the data support, and how this depends on how much data is held out. Each bar shows the fraction
of participants for whom each model was chosen; colours: AR1 (grey), AR1+X1 (light blue, other plane added), AR2 (orange,
second past step added), AR2+X1 (dark blue, both) and AR3 (pink, third past step). Validation: blocked Monte-Carlo
cross-validation (20-step blocks held out at random, 60 repetitions, one mean per participant removed); the chosen model is the
simplest one whose mean held-out error is within one standard error of the best (parsimony rule). First bar of each panel
(``rule''): held-out share set by the dataset's data size, given in the panel title with the number of participants.
Next bars: fixed shares of 20, 50 and 80\% of the blocks. Holding out more data favours simpler models. Top: sagittal
output; bottom: frontal output.}
\label{fig:cv}
\end{figure}

Blocked MCCV with the parsimony rule and the dataset share (50--60\% in OSF walking, 69--84\% in the other walking datasets,
Seethapathi and OSF running, 87--88\% in Maus and Marotta) kept the two-step term in 74 of 85 participants for the sagittal
and 76 of 85 for the frontal output (figure~\ref{fig:cv}). It was dropped mainly in OSF walking (6 of 16 sagittal) and Lee
(3--4 of 13). The cross-plane term was kept in Marotta (6 of 8 sagittal, 8 of 8 frontal), in half of the Seethapathi runners,
in a minority of OSF and de Graaf participants, and never in Lorenz and Maus. With fixed shares the choice moved as expected:
20\% held out kept AR2+X1 in most participants, 80\% dropped the cross-plane term except in Marotta and sent OSF walking (14
of 16 sagittal) and Lee (6 of 13) back to AR1.

\subsection{Drift and measurement noise explain most of the memory}
Differences between trial means were large in most datasets: they explained a median 21\% of the section variance in Lorenz,
48\% in de Graaf, 18\% in OSF walking, 8\% in Marotta, 15\% in Maus and OSF running and 22\% in Seethapathi (three speeds),
but only 4\% in Lee. With one mean per
trial, the two-step gain fell in every dataset (table~\ref{tab:centring}), most in de Graaf, whose participants walked in
seven arm-swing conditions. It stayed positive in every participant. The cross-plane gain fell as well.

\begin{table}[!htbp]
\centering\small
\caption{Median change in held-out error (leave-one-trial-out, shrinkage) with one mean per participant or one mean per
trial. sag / fro output.}
\label{tab:centring}
\setlength{\tabcolsep}{4pt}
\resizebox{\linewidth}{!}{%
\begin{tabular}{@{}l c c c c c c c c@{}}
\toprule
 & \multicolumn{4}{c}{Walking} & \multicolumn{4}{c@{}}{Running} \\
\cmidrule(lr){2-5}\cmidrule(l){6-9}
 & Lorenz & de Graaf & Lee & OSF & Marotta & Maus & Seethapathi & OSF \\
\midrule
Second step, per participant & $-11$ / $-18$\% & $-24$ / $-29$\% & $-4.0$ / $-3.3$\% & $-8.2$ / $-7.0$\% & $-4.8$ / $-10$\% & $-12$ / $-7.9$\% & $-12$ / $-9.7$\% & $-14$ / $-12$\% \\
Second step, per trial       & $-6.2$ / $-12$\% & $-3.5$ / $-6.4$\% & $-2.4$ / $-2.7$\% & $-3.1$ / $-3.1$\% & $-4.0$ / $-7.1$\% & $-7.5$ / $-4.7$\% & $-5.0$ / $-4.7$\% & $-7.8$ / $-7.3$\% \\
Other plane, per participant & $-2.0$ / $-5.4$\% & $-5.8$ / $-7.3$\% & $-1.1$ / $-4.9$\% & $-2.4$ / $-5.8$\% & $-3.4$ / $-6.5$\% & $-0.7$ / $-2.2$\% & $-2.0$ / $-8.6$\% & $-2.4$ / $-7.5$\% \\
Other plane, per trial       & $-0.9$ / $-3.4$\% & $-0.7$ / $-0.9$\% & $-1.0$ / $-3.8$\% & $-1.4$ / $-2.0$\% & $-2.0$ / $-4.5$\% & $-0.2$ / $-0.3$\% & $-1.2$ / $-3.4$\% & $-1.4$ / $-4.6$\% \\
\bottomrule
\end{tabular}}
\end{table}

\begin{figure}[!htbp]
\centering
\includegraphics[width=0.9\linewidth]{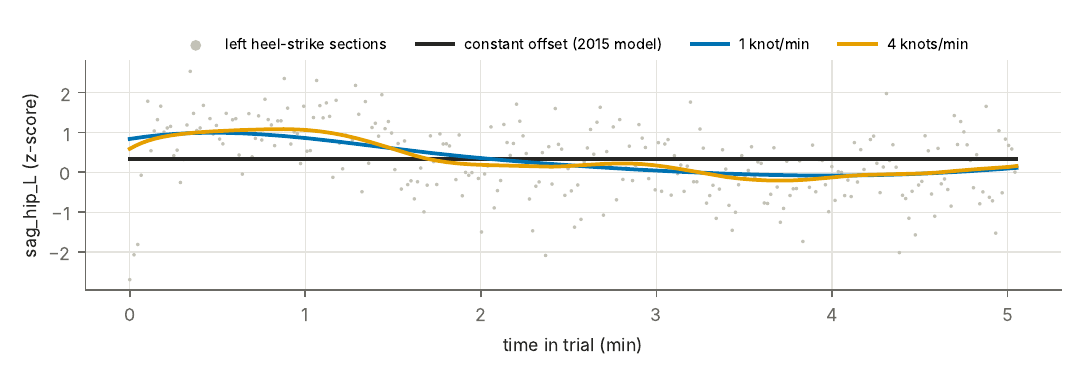}\\[0.3em]
\includegraphics[width=0.95\linewidth]{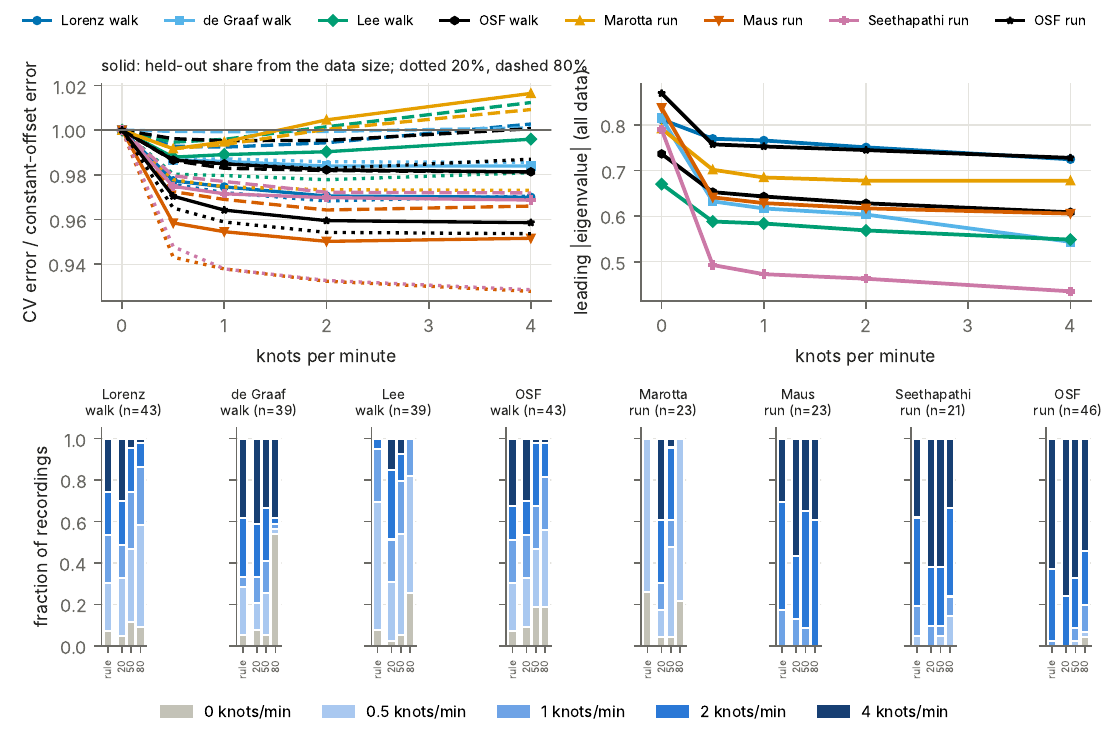}
\caption{A slowly drifting operating point. (\textit{a}) One Lorenz trial: left hip flexion at left heel strike
($z$-score, i.e.\ in units of its standard deviation; grey dots) and the fixed point of the two-step map when the offset
is constant (black, the assumption of earlier studies) or a smooth spline in time with 1 (blue) or 4 (orange) knots per
minute. (\textit{b}) Top left: held-out error with a drifting offset divided by the error with a constant offset, as a
function of knot density (median over recordings; values below 1 mean the drifting offset predicts better). Solid:
held-out share from the data size; dotted and dashed: 20 and 80\% of the blocks held out. Top right: modulus of the
leading eigenvalue of the step map (0: deviations vanish in one step; 1: they persist), fitted on all data; allowing drift
lowers it. Bottom: knot density chosen per recording by the parsimony rule (0 = constant offset), with the held-out
share from the data size (``rule'') and with 20, 50 and 80\% held out; $n$ = recordings.}
\label{fig:drift}
\end{figure}

Within trials the operating point drifted (figure~\ref{fig:drift}a). In the drift test each recording is a unit
(200--1\,700 steps), so its held-out share was 20--26\% in the short walking recordings and up to 84\% in Marotta. The drift
model was kept by the parsimony rule in most recordings: 40 of 43 Lorenz, 37 of 39 de Graaf and 36 of 39 Lee trials, 40 of
43 OSF walking bouts, 17 of 23 Marotta runs and all 23 Maus, 21 Seethapathi and 46 OSF running bouts. The gains were small:
with one knot per minute the error fell by 0.5--4.5\% (figure~\ref{fig:drift}b). With the drifting
offset the leading eigenvalue of the step map moved away from one in every dataset (median 0.81$\to$0.77 Lorenz,
0.81$\to$0.62 de Graaf, 0.67$\to$0.58 Lee, 0.74$\to$0.64 OSF walk, 0.79$\to$0.68 Marotta, 0.84$\to$0.63 Maus,
0.79$\to$0.47 Seethapathi, 0.87$\to$0.75 OSF run at one knot per minute).

\begin{figure}[!htbp]
\centering
\includegraphics[width=\linewidth]{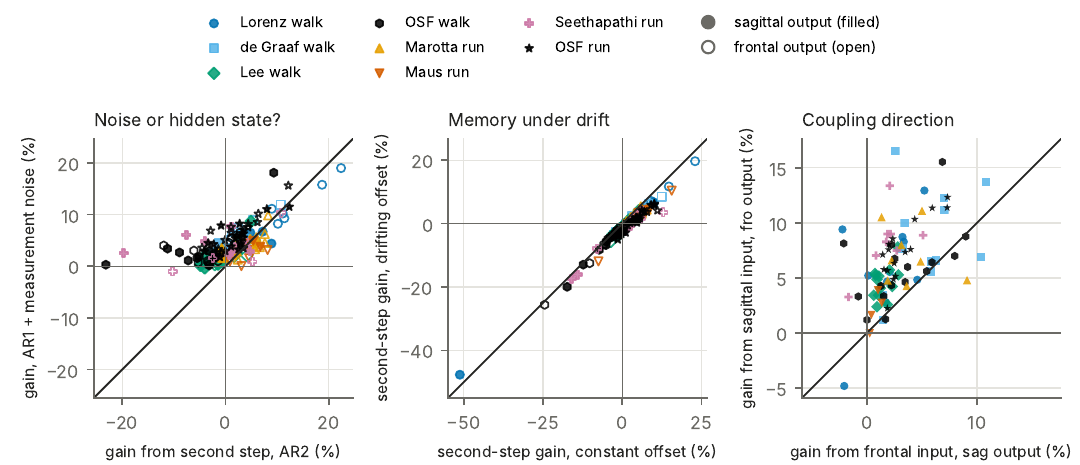}
\caption{Three tests of the two-step memory and of the coupling direction. One point per participant (marker shape and
colour: dataset; filled: sagittal output; open: frontal output); the diagonal marks equal values. Gains are reductions
in held-out error relative to AR1, $100\,(1-e_{\text{model}}/e_{\text{AR1}})$ in \%; positive values mean better
prediction. Left and middle: blocked Monte-Carlo cross-validation with the dataset's held-out share and per-trial means
removed. Left:
gain of AR1 with measurement noise (a first-order state seen through noise) against the gain of AR2 (second past
step); points near or above the diagonal mean that noise explains the apparent memory. Middle: two-step gain with a
drifting offset against the gain with a constant offset per trial; points below the diagonal mean that drift explains
part of the memory. Right: gain from adding the sagittal state to the frontal output against the gain from adding the
frontal state to the sagittal output (leave-one-trial-out validation with population shrinkage, one mean per
participant); points above the diagonal mean stronger sagittal$\to$frontal coupling.}
\label{fig:memory}
\end{figure}

Two tests addressed the memory that remains after per-trial centring (figure~\ref{fig:memory}). Both use participants' own
fits and blocked MCCV with the dataset share (59--88\%). Every candidate keeps the one-step term.

With one map per participant and a drifting offset in each trial, the two-step gain shrank (median, sag / fro): Lorenz
3.7 / 8.4\% $\to$ 1.9 / 5.8\%, de Graaf 0.0 / 2.8\% $\to$ $-1.7$ / 1.1\%, Lee $-1.0$ / $-1.1$\% $\to$ $-2.8$ / $-2.0$\%,
OSF walk $-1.6$ / $-1.0$\% $\to$ $-2.8$ / $-2.1$\%, Marotta 2.3 / 5.2\% $\to$ 1.4 / 3.3\%, Maus 5.8 / 7.0\% $\to$ 3.8 /
3.0\%, Seethapathi 1.0 / 3.4\% $\to$ $-1.5$ / 0.7\%, OSF run 4.3 / 2.7\% $\to$ 0.8 / 0.2\%. Under drift it stayed
positive in most Marotta and Maus runners, in the Lorenz frontal output and in the OSF running sagittal output; in Lee and OSF
walking it was gone, and in Seethapathi it was near zero (figure~\ref{fig:memory}, middle).

AR1 with measurement noise (equation~\eqref{eq:ssm}) predicted better than AR2 in most datasets (figure~\ref{fig:memory},
left; median gain over AR1, AR1+noise vs AR2, sagittal): Lorenz 4.7 vs 3.9\%, de Graaf 2.4 vs $-2.0$\%, Lee 1.3 vs
$-2.6$\%, OSF walk 1.9 vs $-2.5$\%, Marotta 3.2 vs 1.7\%, Maus 4.1 vs 4.8\%, Seethapathi 3.3 vs 0.9\%, OSF run 5.5 vs 2.4\%.
The parsimony rule chose AR1+noise in 43 and 42 of 49 walking participants (sagittal, frontal; AR2 was kept mainly in the
Lorenz frontal output, 6 of 8) and in 30 and 24 of 36 runners, including all 16 OSF runners. AR2 was kept in Maus (3 of 4
for each output) and in the Marotta frontal output (6 of 8), the longest recordings; removing the lumbar angle did not
change this (Marotta frontal 7 of 8, Maus 3 of 4). The estimated measurement noise was 8--12\% of the section variance in the
sagittal plane and 16--22\% in the frontal plane. With half of the blocks held out, AR2 had been kept in 23 of 24 Marotta and
Maus cases and in 10 and 7 of 16 OSF runners; that apparent hidden state depended on the held-out share.

\subsection{The same people keep more memory when they run}
\begin{figure}[!htbp]
\centering
\includegraphics[width=\linewidth]{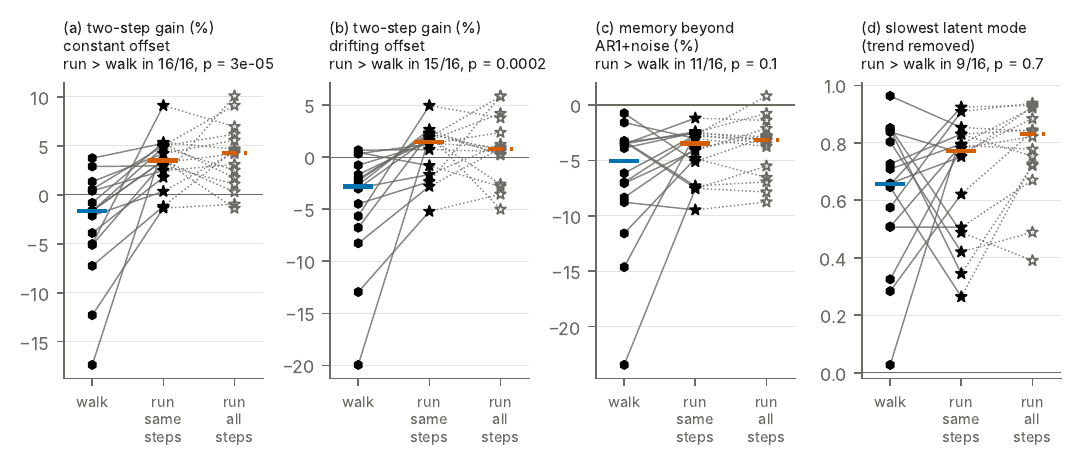}
\caption{Walking (4~km\,h$^{-1}$) and running (9~km\,h$^{-1}$) in the same 16 people (OSF 2026), sagittal output; one
line per participant, short bars: medians. Running is shown cut to the participant's number of walking steps (filled
stars; this pairing gives the test in each title) and with all steps (open stars, dotted). (\textit{a},\textit{b})
Two-step gain, $100\,(1-e_{\text{AR2}}/e_{\text{AR1}})$ in \%, with a constant offset per bout (\textit{a}) and with a
drifting offset (\textit{b}); positive values mean that the second past step improves the prediction. (\textit{c})
Memory beyond AR1 with measurement noise, $100\,(e_{\text{AR1+noise}}-e_{\text{AR2}})/e_{\text{AR1}}$ in \%; negative
values mean the noise model predicts better than AR2. (\textit{d}) Modulus of the slowest latent mode after removing the
minute-scale trend (0 to 1; 1 = very slow). Errors are held-out one-step errors (blocked cross-validation, dataset
held-out share).}
\label{fig:paired}
\end{figure}

In the OSF dataset \cite{krumm2026multimodal,osf2026data} the same 16 people walked and ran on the same treadmill, with the same markers and event rules
(figure~\ref{fig:paired}). Running has a higher cadence, so the same bouts gave about 1\,340 running steps against 830
walking steps. Because the tests depend on data size, running was cut to the same number of steps as each person's walking,
so both gaits had the same held-out share. With a constant offset per bout, the two-step gain (sagittal output) had a median
of $-1.7$\% in walking and 3.5\% in running, and was larger in running in all 16 participants (Wilcoxon signed-rank
$p<10^{-4}$). With a drifting offset it was $-2.8$ vs 1.5\% (15 of 16, $p<10^{-3}$). The memory beyond AR1+noise (AR2 gain
minus AR1+noise gain) was negative in both gaits ($-5.0$ vs $-3.5$\%; larger in running in 11 of 16, $p=0.13$): AR1+noise
was chosen in 15 and 16 of 16 runners. The slowest latent mode did not differ at equal size (0.66 vs 0.77, $p=0.7$). The
fast walk (6~km\,h$^{-1}$) behaved like the slow one. At equal data size, running therefore keeps more step-to-step memory
than walking in the same people, drift does not explain it, and in both gaits AR1 with measurement noise describes it.

\subsection{Spatial coupling: direction and a coupling index}
The sagittal state improved the prediction of the frontal output more than the frontal state improved the sagittal output.
The relative gain of the sagittal$\to$frontal term was larger in 78 of 87 participant-by-gait units (sign test,
$p<10^{-13}$; Lorenz 9/10, de Graaf 9/12, Lee 13/13, OSF walk 13/16, Marotta 7/8, Maus 3/4, Seethapathi 8/8, OSF run 16/16),
and in 74 of 87 once the second step was included
(figure~\ref{fig:memory}, right). The asymmetry is consistent, but both gains are a few percent.

\begin{figure}[!htbp]
\centering
\includegraphics[width=0.95\linewidth]{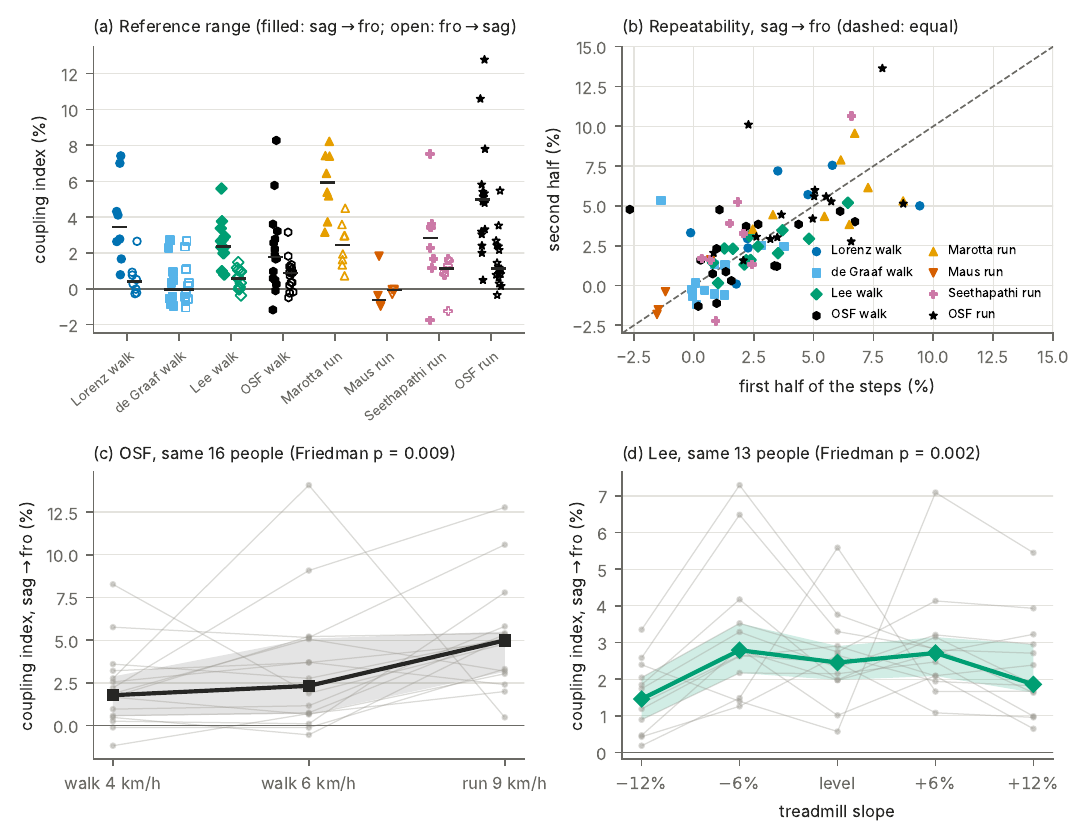}
\caption{Cross-plane (spatial) coupling index, $\mathrm{CI}=100\,(1-e_{\text{AR1+X1}}/e_{\text{AR1}})$ in \%: the
reduction in held-out error when the other plane's state at the previous step is added to the own-plane AR1 map
(blocked cross-validation, dataset held-out share, per-trial means removed, maps shrunk toward the population map).
$\mathrm{CI}=0$: no coupling; $\mathrm{CI}<0$: the extra term only adds estimation error. (\textit{a}) Reference range,
one point per participant; filled: sagittal$\to$frontal (frontal output), open: frontal$\to$sagittal (sagittal output);
bars: medians. (\textit{b}) Repeatability: sagittal$\to$frontal index from the first against the second half of each
participant's steps (dashed: equal values). (\textit{c}) OSF, the same 16 people walking at 4 and 6~km\,h$^{-1}$ and
running at 9~km\,h$^{-1}$ (grey: participants; line and band: median and interquartile range). (\textit{d}) Lee, the same 13 people walking on five
treadmill slopes (grey: participants; line and band: median and interquartile range).}
\label{fig:coupling}
\end{figure}

With shrinkage toward the population map, the sagittal state lowered the frontal held-out error in 73 of 85
participants (median $\mathrm{CI}_{\text{fro}}=2.7$\%, interquartile range 1.0--4.8\%; Wilcoxon $p<10^{-12}$), 2.2\% in
walking and 3.7\% in running (figure~\ref{fig:coupling}\textit{a}). The reverse coupling was smaller
($\mathrm{CI}_{\text{sag}}=0.8$\%, positive in 64 of 85). Without shrinkage, the cross-plane term made the
participant's own map worse in most walkers (positive $\mathrm{CI}_{\text{fro}}$ in 3 of 49) and helped only in the longest
running recordings (Marotta 8 of 8). The index was moderately repeatable (figure~\ref{fig:coupling}\textit{b}): ICC between
the two halves of the steps 0.43 for $\mathrm{CI}_{\text{fro}}$ (within-dataset, all participants; 0.08--0.90 across
datasets), which corresponds to 0.60 for a recording of full length (Spearman--Brown). The smallest detectable change was
1.2--7.3 percentage points. The index changed with the task in the same people. In OSF it rose from 1.8\% (walking
4~km\,h$^{-1}$) and 2.3\% (6~km\,h$^{-1}$) to 5.0\% (running 9~km\,h$^{-1}$; Friedman $p=0.009$; 5.6\% at the walking data
size, larger in 13 of 16, $p=0.003$; figure~\ref{fig:coupling}\textit{c}). In Lee \cite{lee2026adaptive,lee2026data} it varied with slope ($p=0.002$) and was
lowest at the steepest slopes ($-12$\%: 1.5\%, $+12$\%: 1.9\%, level 2.5\%; figure~\ref{fig:coupling}\textit{d}).
$\mathrm{CI}_{\text{sag}}$ did not change with the OSF task ($p=0.65$).

\subsection{Slow dynamics and rank}
\begin{figure}[!htbp]
\centering
\includegraphics[width=0.95\linewidth]{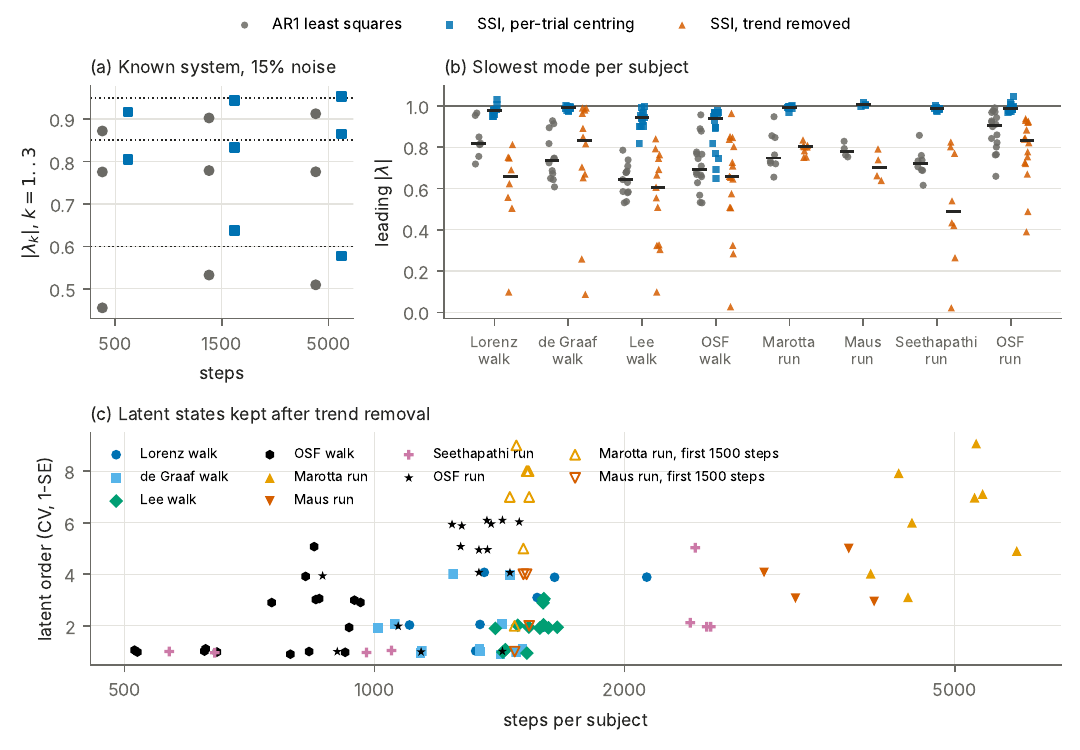}
\caption{Slow dynamics. The modulus $|\lambda|$ of an eigenvalue of the step map (0 to 1) measures how slowly a
deviation decays: it shrinks by the factor $|\lambda|$ per step, so values near 1 are slow modes. (\textit{a}) Known
system with slow modes 0.95 and 0.85 and 15\% measurement noise: the three largest moduli estimated by least squares on
the sections (grey) and by subspace identification (SSI, blue) against data length; dotted: true values. Least squares is
biased toward zero by noise; SSI is not. (\textit{b}) Slowest mode per participant: least squares (AR1), SSI after
per-trial centring, and SSI after removing the minute-scale trend; bars: medians. (\textit{c}) Number of latent states
kept by SSI after trend removal (chosen by blocked cross-validation and the parsimony rule, at least 1) against steps per
participant; open markers: Marotta and Maus cut to 1\,500 steps.}
\label{fig:slow}
\end{figure}

On a synthetic system with known slow modes (0.95 and 0.85) and 15\% measurement noise, subspace identification recovered
them from 500 to 5\,000 steps, whereas least squares gave 0.87--0.91 and 0.78 (figure~\ref{fig:slow}a). In the data, per-trial
centring left latent modes near one in every dataset; they disappeared when the minute-scale trend was removed
(figure~\ref{fig:slow}b), so they were drift. Without the trend, walking kept one to three states (median order 1--2.5)
with a slowest mode of 0.60--0.66 (medians; de Graaf 0.83, mostly frontal, likely Xsens drift): deviations halve within one
to two steps. Marotta and OSF running kept slow modes (median 0.80 and 0.83; a mode above 0.8 in 5 of 8 and 9 of 16
runners), Maus a weaker one (0.70). With each runner cut to 1\,500 steps, the walking size, Marotta still kept a slow mode
(median 0.84, against 0.69 in Lorenz, de Graaf and Lee; Mann--Whitney $p=0.01$), Maus less clearly (0.69). Seethapathi was
closer to walking (0.49). Most slow modes involved both planes.

\label{sec:order0}With half of the blocks held out, cross-validation alone had kept no latent state in most OSF walking
participants (order 0 in 14 of 16), and in the longer datasets cut to the same 830 steps (Lorenz 8 of 8, Lee 12 of 13, Maus 4
of 4, OSF running 12 of 16). Order 0 means $\bm y$ = offset + noise, which contradicts the one-step dependence: with population
shrinkage the one-step map lowered the OSF walking error by 13--22\% (table~\ref{tab:centring}), and the in-sample canonical
correlation between past and future (0.79--0.98) exceeded its white-noise value for the same data size (0.56--0.75) in every
participant. With the dataset share (20\% here) cross-validation alone kept a state in most participants. Order 0 was a
consequence of holding out half of too few data; the order now starts at 1.

\begin{figure}[!htbp]
\centering
\includegraphics[width=0.95\linewidth]{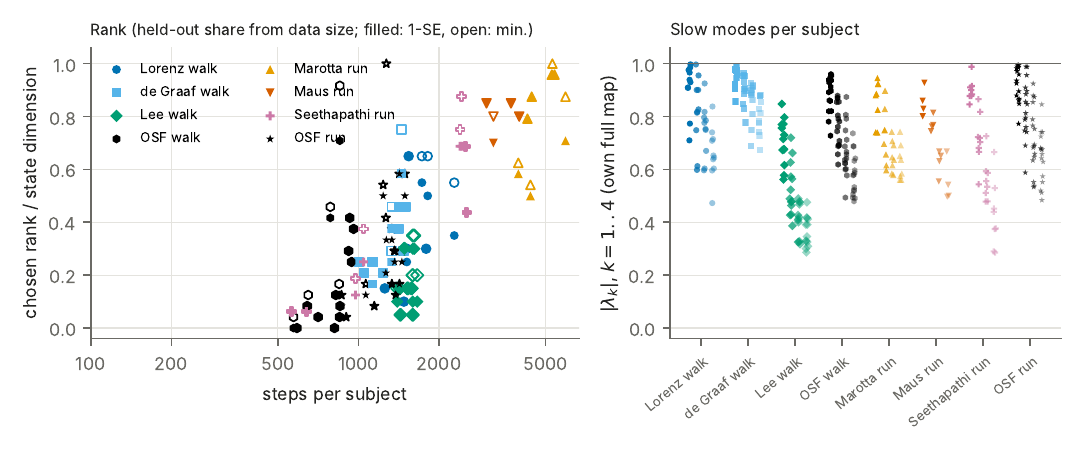}
\caption{Rank and slow modes of the full-state step map (sagittal and frontal planes together). Left: rank chosen by
blocked cross-validation (dataset held-out share), as a fraction of the state dimension, against steps per participant
(log scale); filled: parsimony rule; open: minimum held-out error. A rank below 1 means that only some directions of the
state are predictable from the previous step; more directions are resolved with more data. Right: the four largest
eigenvalue moduli of each participant's full-state map (least squares, one mean per participant removed); values near 1 are slow
modes.}
\label{fig:rank}
\end{figure}

The rank of the full-state map chosen with the dataset share (71--88\%) grew with the number of steps per participant,
from 2.5 of 24 (OSF walk), 3 of 20 (Lee), 6.5 of 20 (Lorenz) and 6.5 of 24 (de Graaf) to 20 of 24 (Marotta) and 16.5 of 20 (Maus) (figure~\ref{fig:rank}, left). The
leading eigenvalue moduli of the participants' own maps had medians of 0.84--0.96, except in Lee (0.72), whose trial means
hardly differ (figure~\ref{fig:rank}, right).

\subsection{The operating point is asymmetric, the step map is not}
\begin{figure}[!htbp]
\centering
\includegraphics[width=0.95\linewidth]{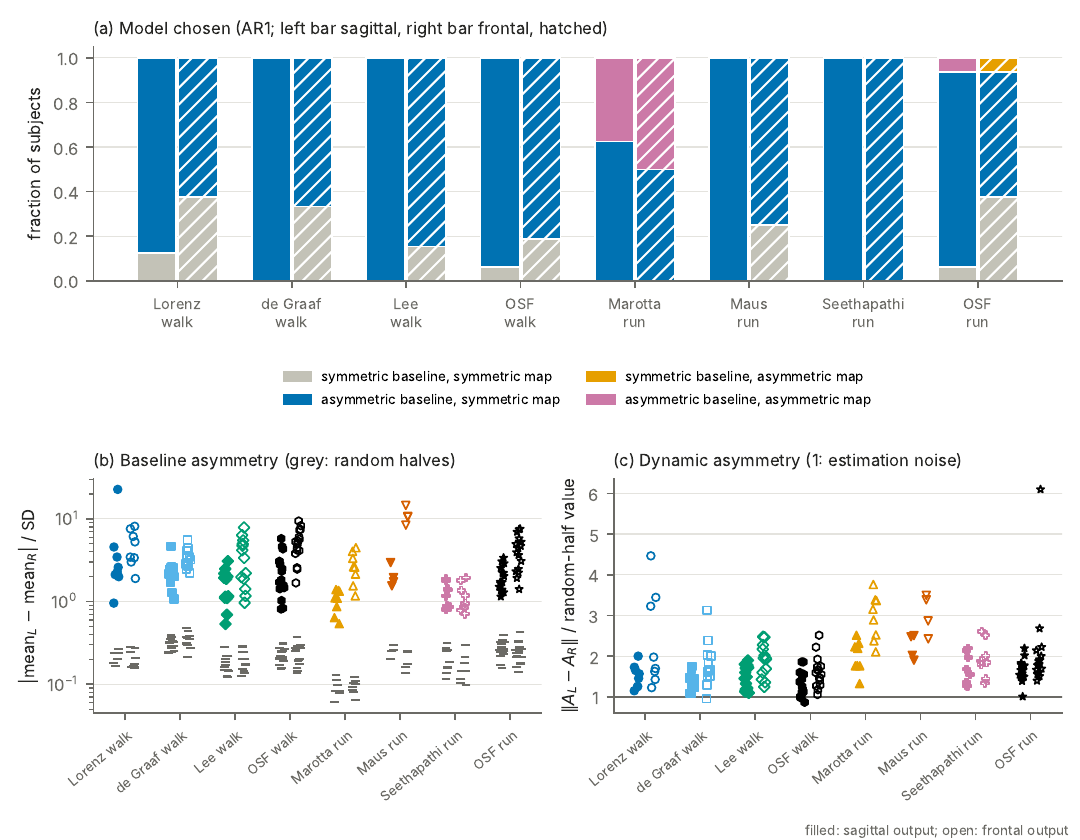}
\caption{Baseline and dynamic symmetry. The baseline is the operating point (mean section of left steps and of
mirrored right steps); the dynamics is the step map. Four models: SS (one offset and one map for both sides), AS
(offset per side, one map), SA (one offset, map per side), AA (both per side). (\textit{a}) Fraction of participants for
whom each model was chosen (AR1 maps, blocked cross-validation, dataset held-out share, parsimony rule); left bar of each
pair: sagittal output; right, hatched: frontal output. (\textit{b}) Left--right difference of the mean sections in units
of the step-to-step standard deviation (log scale; root mean square over variables, median over trials); grey dashes: the
same statistic for two random halves of the left steps, i.e.\ the estimation noise. (\textit{c}) Difference between the
left and right maps, $\|A_L-A_R\|$, divided by the difference between two maps fitted on random halves of the steps; 1
means the maps differ no more than estimation noise. Filled: sagittal; open: frontal output.}
\label{fig:symmetry}
\end{figure}

Symmetry has two parts: the \emph{baseline} (the mean section of left and of mirrored right steps) and the
\emph{dynamics} (the step map). We compared four models with offsets per trial fitted together with the map: one offset
and one map (SS), one offset per side and one map (AS), one offset and one map per side (SA), and one offset and one map per
side (AA). The baseline was asymmetric: the left--right difference of the mean sections was 1.8 (sagittal) and 3.6
(frontal) step-to-step standard deviations (medians), against 0.2 for random halves, and an asymmetric baseline was chosen
in 82 of 85 participants for the sagittal and 65 of 85 for the frontal output (figure~\ref{fig:symmetry}). The dynamics were
symmetric within what the data resolve: the left and right maps differed by 1.5--1.8 times the estimation noise, but
separate maps were chosen in only 4 and 5 of 85 participants, mostly on the bends of the Marotta track. An unmodelled
baseline asymmetry also looks like memory: with one offset for both sides, the two-step gain rose from 10.4 to 14.3\%
(sagittal, medians; larger in 82 of 85 participants), because the state two steps back carries the side. All other analyses
therefore remove the operating point per side.

\section{Discussion}

In this paper, we focused on temporal and spatial coupling in the step-to-step dynamics of human walking and running. We
fitted linear maps between Poincar\'e sections of seven public datasets and used cross-validation to ask how wrong, and
how useful, three common simplifications are: a Markov state, a constant operating point and separate planes. The section
state looks non-Markov at first: the state two steps back improves the prediction of the next step. Most of this apparent
memory, however, comes from a drifting operating point and from measurement noise. Running keeps more of it than walking.
The planes are coupled, but weakly. And how much data is held out for validation decides much of what one concludes.

\subsection{Temporal coupling and a slowly moving limit cycle}
Most of the memory in walking disappears once the operating point is allowed to differ
between trials and to drift within them, and most of the rest is explained by measurement noise on a first-order (AR1) map. The
walking dataset with the most stable operating point (Lee: trial means explain 4\% of the variance) has the least memory
(3--4\%), and in Lee and OSF walking nothing is left once the operating point may drift. A drifting
operating point is expected from how humans regulate steps. Step length follows slow speed fluctuations \cite{collins2013two},
deviations along task-equivalent directions persist \cite{dingwell2010humans,dingwell2015identifying}, and a model of
task-level regulation is only semistable \cite{patil2024model}. Slow components of this kind also produce the long-range
correlations of stride intervals \cite{hausdorff1995walking,dingwell2010reinterpreting,terrier2026persistent}. The estimated
measurement noise, 8--22\% of the section variance and larger in the frontal plane, is consistent with the known accuracy of
marker and inertial joint angles outside the sagittal plane \cite{leardini2005human,zhang2013concurrent,robertlachaine2017validation}.
For walking, a first-order map with a drifting operating point and noisy observations is therefore an adequate description at
the section.

Relaxing the constant operating point does not remove the
limit-cycle picture. The operating point moved smoothly and slowly compared with the step-to-step dynamics, while the
step map itself stayed the same within a recording: a map that differed between the early and the late half of a trial
was chosen in 1 of 172 participant-by-plane cases (ESM S8). Gait therefore still behaves like a stable limit cycle. What
changes is its position: a slowly drifting operating point, around which deviations decay through a fixed step map.
Separating the two time scales is what removes most of the apparent memory.

Running keeps more step-to-step memory than walking, across datasets
and within the same people at equal data size (15--16 of 16); laboratory, sensors and event rules cannot explain this. Most of it
is still described by AR1 with measurement noise. A further state is supported only in the longest recordings
(Maus; the Marotta frontal output), with or without the trunk angle. This agrees with Maus et al.\
\cite{maus2015constructing}, who found that a spring-mass state is not enough to predict the next step of running and that
swing-leg states are needed, and with the step-to-step corrections of the centre of mass observed in running
\cite{seethapathi2019step}. Candidates for such a state are the centre-of-mass state, the swing-leg state and leg stiffness,
which are not part of our joint-angle state at foot strike. With half of the data held out, the two-step model won in almost
every long running recording; the conclusion ``running needs a hidden state'' therefore depends on the validation design and
needs longer recordings, ideally of the same people walking and running.

Once drift is separated, walking has no slow step-scale mode, while running has one (Marotta and
OSF running). Its detection needs data: at the walking data size the same OSF runners did not differ from walking. Least-squares
eigenvalues, the usual estimate of Floquet multipliers, were raised by drift (by 0.12--0.29) and lowered by noise (by up to
0.34), so their net bias depends on the recording. The latent order also depends on the validation design: with half held out, short recordings kept no state at all, which
contradicts the one-step dependence; a share that follows the data size avoids this, and the order should start at 1.
Slow-mode comparisons should be made at equal data size.

\subsection{Spatial coupling and symmetry}
The planes are coupled from one step to the next. The coupling is
found in most participants of every dataset, runs mainly from the sagittal to the frontal plane, and is a robust result of
this study. It is also small: the other plane lowers the prediction error by a few percent, much less than the drift and
the two-step term. For healthy gait this supports the common practice of treating the planes separately: planar models of
walking and running in the sagittal plane \cite{mcgeer1990passive,garcia1998simplest,geyer2006compliant}, and analyses of
lateral stability and foot placement that treat the frontal plane on its own
\cite{kuo1999stabilization,bauby2000active,hof2005condition,dingwell2019humans}. The direction is consistent with active
lateral control through foot placement \cite{kuo1999stabilization,bauby2000active,oconnor2009direction,wang2014stepping,bruijn2018control}
and with shared muscular contributions to fore--aft and lateral foot placement \cite{roelker2019muscle}; the gains are too
small to show the mechanism.

A small but reliable coupling may still be useful as a measure of gait quality. Many gait disorders couple the planes:
after stroke, a reduced sagittal swing is compensated by hip hiking and circumduction in the frontal plane
\cite{kerrigan2000hip,chen2005gait}, and the control of lateral foot placement is impaired \cite{dean2015foot}; in older
adults, step-width variability is associated with falls \cite{brach2005too}. Gait variability is already used as a
clinical marker \cite{hausdorff2005gait,kuo2010dynamic}, but it does not say whether deviations in one plane are corrected
in the other. The coupling index does. In our healthy data it was small, positive, moderately repeatable and sensitive to
speed, gait and slope; a gait whose index lies far outside the healthy range, or changes during rehabilitation, would
point to a change in how the planes are controlled together. Two requirements follow from our results. The index needs a
normative population map, since individual maps without shrinkage are dominated by estimation error, and it needs long
recordings: the smallest detectable change was of the order of the index itself at the walking data sizes here. Whether
the index separates patients from controls, or tracks recovery, has to be tested in clinical data.

Separating the baseline from the dynamics refines the earlier finding that a single mirrored map
describes walking well enough \cite{ankarali2015walking}. The step map is symmetric within what several thousand steps
resolve, in walking and running, while the operating point is clearly asymmetric, as expected from the small asymmetries
of healthy gait \cite{sadeghi2000symmetry} and from marker placement. A side-specific operating point has to be removed
before memory is assessed; otherwise the alternating left--right difference passes for two-step memory. The contrast also
shows why the dynamics are the more robust measure. A constant error in marker placement or calibration shifts the
operating point and can make it look asymmetric. It does not change the eigenvalues of the step map, which are invariant
to any constant offset and to any constant invertible linear transform of the measured state. A side-specific linear
error can make the single-step maps $A_L$ and $A_R$ differ, but not the eigenvalues of the stride map $A_RA_L$. The
dynamics remain sensitive to measurement noise and to slow drift, which the noise and drift models address.

\subsection{What the models can and cannot tell}
Most memory and coupling terms changed the held-out error by
a few percent. A term kept by the parsimony rule is detectable in these data. It is not, for that reason, physiologically
meaningful, and we do not read it as a control mechanism. Four observations support this.
First, detection depends on the amount of data. With thousands of steps, cross-validation resolves gains of 1--2\%. The
same terms were dropped with less data or a larger held-out share, and the rank and the latent order grew with the data.
Cross-validation picks the best predictor that the data support, not the structure of the controller
\cite{shao1993linear,arlot2010survey}.
Second, most step-to-step variance is not predictable. If the sections were a first-order state seen through noise, no
predictor could do better than 64--85\% of the section variance (ESM S8). The memory terms act on the small remainder.
Third, terms of this size arise from the analysis itself. Drift, measurement noise and an unmodelled baseline asymmetry
each produced two-step gains of several percent that disappeared once they were modelled. A remaining term of a few percent
may have causes of the same kind that we did not model, such as event-timing jitter, nonlinearity of the map or the choice
of section.
Fourth, prediction error is not a behavioural outcome. Whether a term matters for gait would show in its effect on
stability or on the response to a perturbation, which unperturbed data cannot test. Experiments that move the state by a
known amount are the way to ask whether a weak term reflects control \cite{cowan2014feedback}.
We therefore treat the small terms as properties of the data and of the model class. This does not prevent a small
term from being a useful marker, if it is repeatable and changes with the condition of interest; the coupling index
above is an example. The findings are the large and
consistent effects: apparent memory from drift and noise, the role of the validation design, the asymmetric baseline with a
symmetric map, and more memory in running than in walking in the same people.

Floquet multipliers, Lyapunov exponents and fluctuation analyses are
computed from section data or embedded states that assume a fixed operating point and, implicitly, a Markov state
\cite{hurmuzlu1994measurement,dingwell2007differences,bruijn2014assessing}. Our results show three ways in which this matters.
First, removing one mean per participant leaves slow components in the data, which inflate both apparent memory and the
leading eigenvalue; allowing drift lowered the leading eigenvalue in every dataset. Second, measurement noise produces
apparent memory, so the order of a fitted model is not a property of the gait alone. Third, how much data is held out for
validation decides which terms are kept. We suggest that studies of gait dynamics report how the operating point was
removed, test the memory order by cross-validation, and report the validation design.

The validation design decided which terms were kept. Holding out more data favoured simpler models,
as expected from theory \cite{shao1993linear,arlot2010survey}; the two-step term survived every design, while the cross-plane
term survived only with long recordings. A fixed held-out share is not neutral: with few data it leaves the largest model
underfitted and favours the empty model, and with many data it is less strict than the data allow. A share that grows with
the data, after enough is kept for fitting, removed the spurious ``no dynamics'' result in short walking recordings and the
spurious ``hidden state'' in most running recordings. Shrinkage toward the population map was the most reliable way to fit individual
maps, and the rank depended on the amount of data per participant.

The methods use only kinematics at the sections and public data, and they apply to any
rhythmic system with alternating, nearly symmetric steps: bipedal walking and running, quadrupedal walking and trotting,
and hexapedal tripod gaits. In legged robots, the same tests can serve as diagnostics. A drifting operating point, a
coupling between the planes or a side-specific map would point to calibration errors, wear or a steering bias, and only
internal sensors are needed. On the biological side, models of locomotion can be decomposed into the mechanics (plant) and
the neural feedback (controller) \cite{roth2014comparative}. A less wrong model of the plant gives a better understanding of
the controller, and vice versa \cite{cowan2007critical,cowan2014feedback}. Separating a slowly moving operating point and
measurement noise from the fast step map is a step in this direction: what remains is closer to the closed-loop dynamics
that the controller shapes.

\subsection{Limitations and outlook}
The datasets differ in sensors, state variables, event detection and conditions, and the running
datasets are small (8, 4, 8 and 16 participants); the OSF bouts are short (3 min), so its walking--running
comparison is limited to about 830 steps per person, and its running sides were assigned from hip flexion. Joint angles in Seethapathi come from marker clusters without a static trial. Treadmill and overground walking differ in variability \cite{dingwell2001local,hollman2016comparison};
only one dataset is overground, and it includes the bends of a running track. The maps are linear and AR1+noise has as
many latent states as outputs. The measurement noise in the model also absorbs timing jitter of the detected events, which
maps into the section state through the state velocity \cite{mo2018accuracy}. All results use heel-strike sections; other phases of the stride changed
predictability and the size of the memory, but not the conclusions (ESM S6). The coupling direction was assessed with
relative gains of outputs of different size.

A latent model with more states than outputs \cite{vanoverschee1996subspace} would estimate the dimension
of a possible further state in running and could be compared with the slow-dynamics dimension of Revzen \& Guckenheimer
\cite{revzen2012finding}. Floquet multipliers computed on the same data with and without drift and noise models would quantify
the bias of standard stability measures. Long overground recordings of walking would show whether the walking result holds off
the treadmill.

\section{Conclusion}
The heel-strike state of human gait appears non-Markov, and its planes appear coupled. Most of the apparent memory is a
slowly drifting operating point and measurement noise on a first-order map; gait remains limit-cycle-like around an
operating point that moves slowly. Running keeps more memory than walking, also within the same people at equal data
size, and a further state appears only in the longest recordings. The sagittal and frontal planes are coupled, mainly
sagittal to frontal, but weakly; a coupling index is repeatable and task dependent, and is a candidate measure of gait
quality for clinical testing. These conclusions, and any stability measure built on step-to-step maps, depend on how the
operating point is modelled and how much data is held out for validation. Like bilateral
symmetry in our earlier work, the Markov and planar assumptions are wrong in a strict sense. Around a slowly moving
operating point, they are useful enough.

\section*{Ethics}
This study is a secondary analysis of publicly available, de-identified data. No new data were collected. Each original
study obtained ethical approval and informed consent, as reported in its publication
\cite{lorenz2024comprehensive,degraaf2019influence,lee2026adaptive,marotta2021towards,maus2015constructing,seethapathi2019step,krumm2026multimodal,moore2015elaborate}.

\section*{Funding}
This work was partially supported by the Health Institutes of T\"urkiye (T\"US\.{E}B) under project no.\ 40908.

\section*{Acknowledgements}
I thank Noah J.\ Cowan, whose mentorship shaped my perspective on the dynamics of humans and animals. I thank the authors of the datasets used here for making their data public.

\bibliographystyle{rsif}
\bibliography{refs}

\end{document}